# The True Role of Active Communicators:
# An Empirical Study of Jazz Core Developers

Sherlock A. Licorish and Stephen G. MacDonell

*SERL, AUT University*
*Auckland, 1142, New Zealand*
*slicoris@aut.ac.nz, smacdone@aut.ac.nz*

## ABSTRACT

**Context:** Interest in software engineering (SE) methodologies and tools has been complemented in recent years by research efforts oriented towards understanding the human processes involved in software development. This shift has been imperative given reports of inadequately performing teams and the consequent growing emphasis on individuals and team relations in contemporary SE methods. **Objective:** While software repositories have frequently been studied with a view to explaining such human processes, research has tended to use primarily quantitative analysis approaches. There is concern, however, that such approaches can provide only a partial picture of the software process. Given the way human behavior is nuanced within psychological and social contexts, it has been asserted that a full understanding may only be achieved through deeper contextual enquiries. **Method:** We have followed such an approach and have applied data mining, SNA, psycholinguistic analysis and directed content analysis (CA) to study the way core developers at IBM Rational Jazz contribute their social and intellectual capital, and have compared the attitudes, interactions and activities of these members to those of their less active counterparts. **Results:** Among our results, we uncovered that Jazz's core developers worked across multiple roles, and were crucial to their teams' organizational, intra-personal and inter-personal processes. Additionally, although these individuals were highly task- and achievement-focused, they were also largely responsible for maintaining positive team atmosphere. Further, we uncovered that, as a group, Jazz developers spent a large amount of time providing context awareness in support of their colleagues. **Conclusion:** Our results suggest that high-performing distributed agile teams rely on both individual and collective efforts, as well as organizational environments that promote informal and organic work structures.

**Categories and Subject Descriptors**
D.2.9 [**Software Engineering**]: Management – Productivity, Programming Teams.

**General Terms** Human Factors, Management.

**Keywords** Software Development, Core Developers, Roles, Psycholinguistics, Content Analysis, Behaviors, Jazz.

## 1. INTRODUCTION

Concerns over software systems' adequacy, project success rates and the adoption of appropriate software process models have been ubiquitous and longstanding [1-2]. Despite many recommendations in relation to adopting various software methodologies and tools [3-4], there remain questions over the outcomes of software development projects [5-6]. Previous evidence suggests that people factors manifest in communication and behavioral issues, and these underscore the causes of inadequately performing software teams [7-9]. Thus, studying these issues should provide fruitful avenues for researchers to better understand the software process and to offer recommendations for process improvements.

This position is supported by the considerable body of research dedicated to studying human issues [10-11] and, particularly in recent years, there has been growing emphasis on studying the communication and coordination practices of software engineering teams [12-15]. These works have examined software artifacts to inform our understanding of the way different software teams co-exist in order to provide software solutions. Software repositories have also received noteworthy attention in this regard [16-17]; however, these studies have tended to employ primarily quantitative, mathematically-based analysis techniques (e.g. SNA). While these techniques do enable the detection of certain patterns, and so provide a partial understanding of software teams' behavioral processes, there are limitations on the effectiveness of these approaches in informing our understanding of the deeper psychosocial nature of team dynamics [18].

Given the way contemporary software development is driven, with particular emphasis on individuals and relationships as compared to processes and tools [19], it seems pertinent to supplement such mathematical analyses with more contextual examinations if we are to more fully comprehend the nature of agile teams. For example, previous work has uncovered that just a few individuals dominate project communications [11]. However, questions related to *how* and *why* this phenomenon exists, why core group members become 'knowledge hubs', the reasons for these members' extraordinary presence, and understanding the actual roles (both formal and informal) that core developers occupy in their teams, have not been answered. Such explorations could provide explanations for the peculiarities of agile group dynamics, may inform appropriate team configurations, and may enable the early identification of 'software gems' – exceptional practitioners in terms of both task and team performance.

To this end we have followed a systematic analysis approach using software repository data to study the way core developers enact various roles and behaviors during their projects. We have examined practitioners' messages and logs of change histories while these individuals were working on various forms of software tasks. Our analysis was conducted in two main phases. In the first phase the IBM Rational Jazz repository was mined and multiple project areas were explored using SNA and standard statistical techniques. This phase allowed us to select our specific project cases and to detect patterns around core developers. This informed the design of the second phase, which involved the use of deeper linguistic analysis techniques and directed CA. The findings from these investigations are reported here.

In the next section (Section 2) we present our theoretical background and survey related works. This leads to our specific research questions outlined in Section 3. We then describe our research methods and measures in Section 4, introducing our techniques and procedures in this section. In Section 5 we present our results and analysis. Section 6 then discusses our findings and outlines the implications of our results, and in Section 7 we consider our study's limitations. Finally, in Section 8 we draw conclusions and outline further research directions.

## 2. RELATED WORK

It is widely recognized that software artifacts and software history data are useful sources of interaction evidence [20]. Findings from works examining such data are especially valid if the artifacts under investigation constitute the primary means of interaction and evidence of team processes during software development. Thus, previous researchers have exploited process artifacts such as electronic messages, change requests histories, bug logs and blogs to provide unique perspectives on activities occurring during the software development process [12, 21].

Open source software (OSS) repositories and archives recording software developers' textual communication activities have particularly supported enquiries aimed at understanding software practitioners' social behaviors [22-23]. For instance, Abreu and Premraj [17] observed the ECLIPSE mailing list and found that increases in communication coincided with a high number of bug-introducing changes, and developers communicated most frequently at release and integration time. Additionally, these authors discovered that the number of messages increased after software check-ins, and communication levels of the entire project team (rather than individual developers) are indicative of the number of bug-introducing changes.

Bird et al. [24] employed clustering algorithms to study CVS and mailing lists to unearth coordination and communication activities of Apache developers. Evidence confirmed that the more software development an individual does the more coordination and controlling activities they must undertake, and the volume of messages an individual sends is no indication of an individual's position in the group. These findings are in contrast to those of Cataldo et al. [12], whose SNA studies found that frequent communicators contributed the most during software development. The findings of Bird et al. [24] may be used to assess periods of team productivity and for identifying the most active developers. That said, the contention in their study that the regular communicators are not the 'main team members' is quite revealing, and signals the need for further research. In fact, Cataldo et al. [12]'s findings have been supported by other similar studies. Using the GTK and Evolution OSS projects Shihab et al. [25] also established that only a small number of developers participated in internet relay chat (IRC) meetings. Similarly Shihab et al. [11] found communication activity to be correlated with software development activity when studying the OSS GNOME project, where what was communicated was reflected in source code changes. Shihab et al. [11] observed that the most productive developers contributed 60% of the project communication and their interaction levels remained stable over the project duration when compared to lesser contributing participants.

In commercial settings (or closed source software (CSS)) the IBM Rational Jazz repository has been used in the study of software practitioners' interactions and communications largely from a SNA perspective [16, 26-27], offering contradictory findings to those drawn from the OSS-based body of work. Contrary to the findings reported by Shihab et al. [11, 25], Nguyen et al. [16] uncovered that about 75% of Jazz's core team members actively participated in the project's communication network. Additionally, these authors found Jazz project teams to have very inter-connected social networks, requiring few brokers to bridge communication gaps. These findings may be reflective of software practitioners' disposition in commercial settings, where team members' motivation to contribute their knowledge is likely to be driven by greater rewards when compared to those received in OSS environments.

Moreover, previous research has cautioned about the inferences and generalizations drawn based on analyses of the regularly extracted OSS repositories that are often used to study process issues [24, 28]. Particular challenges that arise when using OSS repositories relate to the reliability and validity of the data available. Research evidence has reported poor data quality in some repositories of OSS projects [24, 28]. For instance, in their study of the Apache mailing list, Bird et al. [24] found it difficult to uniquely identify developers' records due to the volume of email addresses and aliases these individuals used. Further issues may also be encountered when studying OSS repositories because anyone is able to post messages and report bugs to such mailing lists, whether or not those individuals are contributing to the project [29].

On the other hand, commercial software organizations seldom make their code history or project data publicly available. Additionally, previous work examining repositories such as those comprising IBM Rational Jazz [16] and Microsoft [30] datasets have tended to employ mathematically-based analysis techniques. While these approaches have provided much needed understandings related to the way software practitioners work, there still remain questions over how practitioners co-exist during agile projects, and especially during distributed agile developments. We believe that it is timely to examine the contextual interactions and engagements of agile practitioners if we are to more fully comprehend the nature of agile teams. This is particularly necessary given the agile stance of favoring individuals and interaction, against an emphasis that focuses on software variations and changes in artifacts such as requirements documents and code [2].

## 3. RESEARCH QUESTIONS

Several of the studies introduced in Section 2 uncovered that only a small number of team members tend to contribute to a project's knowledge base [11, 25], and that software developers' communication and coordination activities are directly related to their involvement in software tasks [24]. As noted above, questions related to *how* and *why* core group members become 'knowledge hubs', the reasons for these members' extraordinary presence, and understanding the actual roles (both formal and informal) that core developers occupy in their teams, have not been answered. Such answers could provide explanations for the nature (and peculiarities) of agile group dynamics. Knowledge and awareness of the way the most active agile practitioners contribute their social and intellectual capital could help project leaders to identify exceptional software practitioners, and inform the process of assembling high performing and cohesive teams. Such findings could also inform the use of specific organizational arrangements and team configurations in support of high performers. Furthermore, the output of these explorations may lead to new requirements for collaboration and process support tools. We therefore examine a CSS repository using exploratory, linguistic and contextual analysis techniques in order to answer the following questions:

   Q1. What are the core developers' enacted roles in their teams, and how are these roles occupied?

   Q2. How do core developers' behaviors and attitudes differ from those of other software practitioners?

## 4. METHOD AND MEASURES

During a large-scale project utilizing a multiple case study design to investigate team evolution and dynamics within the IBM Rational Jazz repository we observed that a few individuals dominated project interactions (see [31] where we examine the effect of project environment (project type, people involved) on team behaviors). These variations were detected while examining sociograms we created from practitioners' messages during initial SNA explorations (see Figure 1 for an illustration). In this present work we employ an embedded multi-case design [32] to systematically examine core developers' enacted roles and behaviors when undertaking software development tasks. Our goal is to understand how they contribute to team dynamics. An embedded case study approach is appropriate for understanding complex human processes by relating them to their context [32], as is the focus of the work under consideration here. During our study, mining methods were used for data collection (see subsection 4.2) and the extracted data were then analyzed using statistical techniques, linguistic analysis tools (see subsection 4.3) and directed CA (see subsection 4.4). Below we first provide a description of the repository that was used as the data source in our study (subsection 4.1) and then we elaborate on the techniques and procedures utilized during our research.

### 4.1 Study Repository

The repository examined in this work is a specific release of IBM Rational Jazz (based on the IBM[R] Rational[R] Team Concert[TM] (RTC)[1]). Jazz, created by IBM, is a fully functional environment for developing software and managing the entire software development process, including project management, project communication and coding [33]. The software includes features for work planning and traceability, software builds, code analysis, bug tracking and version control in one system [34]. Changes to source code in the Jazz environment are only allowed as a consequence of tasks created beforehand, such as a bug report, a new feature request or a request to enhance an existing feature. Features and artifacts are tracked using work items (WIs), and a WI represents a single task which is one of the three task types just mentioned. Team member communication and interaction around WIs are captured by Jazz's comment or message functionality. In fact, during development at IBM, project communication (the content explored in this study) was and is actually enforced through the use of Jazz itself [16].

IBM permitted us to study an instance of the Jazz repository comprising a large amount of process data from development and management activities across the USA, Canada and Europe. This instance (release 1.0.1) included numerous projects, many of which have since been released and are commercially available (e.g., RTC), with specific teams responsible for each project (although team members also work across projects). Each team has multiple individual roles, with a project leader responsible for the management and coordination of the activities undertaken by the team. Project leaders report progress to a project management committee, which formulates and oversees project goals. Jazz teams use the Eclipse-way methodology for guiding the software development process [33]. This methodology outlines iteration cycles that are six weeks in duration, comprising planning, development and stabilizing phases. Builds are executed after project iterations; also called project milestones. All information for the software process (project management, tracking and planning, project coordination and communication, and software building/coding) is stored in a server repository, which is accessible through a web-based or Eclipse-based (RTC) client interface [26].

---

[1] IBM, the IBM logo, ibm.com, and Rational are trademarks or registered trademarks of International Business Machines Corporation in the United States, other countries, or both.

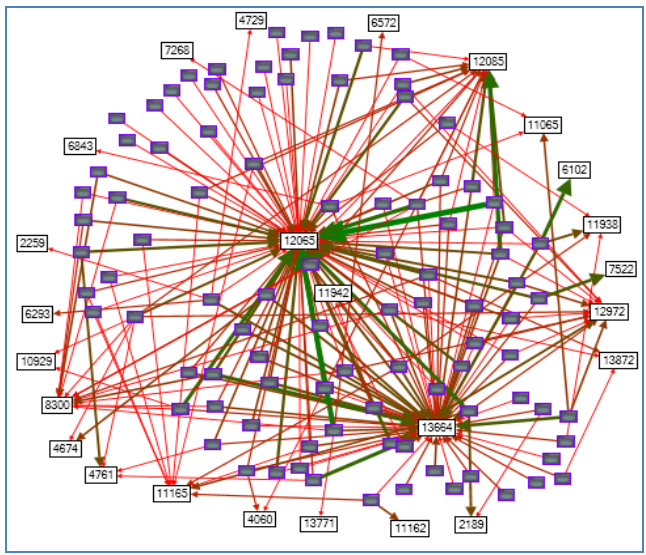

**Figure 1. Directed network graph for a sample Jazz project showing highly dense network segments for practitioners '12065' and '13664'**

## 4.2 Data Pre-processing and Metrics Definition

### 4.2.1 Data Extraction

We created a Java program to leverage the IBM Rational Jazz Client API to extract team information and development and communication artifacts from the Jazz repository. These included:

- Work Item (or Software Task) history logs – in Jazz each software task is represented as a WI (see subsection 4.1), and a history log is maintained for each WI.

- Project Workspace or Project Area – each Jazz team is assigned a workspace. The workspace, also called the project or team area, contains all the artifacts belonging to that team.

- Contributors and Teams – a contributor is a practitioner contributing to one or more software features, multiple contributors form teams.

- Comments or Messages – communication around WIs is captured by Jazz's comment functionality. Messages ranged from as short as one word (such as 'thanks'), to up to 1055 words representing multiple pages of communication. These are arranged by date sequentially for each WI, similar to messages on a bulletin board.

We extracted 36,672 resolved WIs and their associated history logs (from the various project areas in the repository) created between June 2005 and June 2008. These work items belonged to 94 Project Workspaces that each comprised more than 25 WIs. We were therefore confident that we had access to a sufficient volume of data to support the planned investigations. The project workspaces consisted of 474 active contributors belonging to eight different roles. For the 94 project areas, comments (or messages) – our primary data source – were also extracted, totaling 116,020. The data extracted from Jazz were imported into a relational database management system to facilitate efficient data management.

In line with the multi-case study design used in our overall work (mentioned above), purposive sampling was then applied in the case selection process [32]. The goal was to select a range of cases that represented the scope and breadth of the various project areas in the repository, for example: documentation, user experience, development or coding, and project management-based activities. We therefore selected all tasks undertaken by ten of the 94 project teams (shown in Table 3). The project areas selected represented both information-rich and information-rare cases in terms of messages. These cases also represented projects that were long, medium and short in duration. As the data were analyzed, it became clear that the cases selected were indeed representative of those in the repository for team members' messages and contribution to task changes, as we reached data saturation [35] after analyzing the third of the ten project cases.

### 4.2.2 Procedure and Metrics

In SNA, mathematically defined concepts such as cohesion, equivalence, power and brokerage are used to explain the characteristics of the network and its actors [36]. One of the most important characteristic for SNA is cohesion, measured by 'density' and 'centrality'. Density provides an overall measurement of the connectedness of a network [36] and was used here in order to select Jazz's 'core developers'. Density varies between 0 and 1, so that a contributor who communicated on all tasks would have a density of 1, while one with no interaction would have a density of 0. Individuals involved in highly dense communication network segments have been shown to dominate coordination and collective action [37], and are seen as most important to their teams [38]. We created a baseline using a similar approach to that used by Crowston et al. [39], and selected all practitioners whose density measure was ≥ 0.33 (i.e., they communicated on a third or more of their teams' project tasks), and labeled them as top contributors or 'core developers'.

We first extracted these practitioners' formal team role and responsibility information as recorded in the repository, and task change history data were analyzed to assess the involvement of these core developers in software tasks. We then discerned the *enacted roles* (roles that are actually acted out during the project) of these members by studying the nature of their interactions through the messages they communicated (e.g., whether these members instructed others, provided coordination expertise, and so on). The behaviors of these and other members were studied using linguistic analysis (see subsection 4.3) and were triangulated with deeper contextual analysis (see subsection 4.4).

## 4.3 Linguistic Analysis Techniques

Previous research has identified that individual linguistic style is quite stable over time and that text analysis programs are able to accurately link language characteristics to behavioral traits (see [40], for example). We employed the Linguistic Inquiry and Word Count (LIWC) software tool in our analysis. The LIWC is a software tool created after four decades of research using data collected across the USA, Canada and New Zealand [41]. Data collected in creating the LIWC tool included all forms of writing and normal conversations. This tool captures over 86% of the words used during communication (around 4500 words). Words are counted and grouped against specific types, such as negative emotion, social words, and so on (see Table 1). Written text is submitted as input to the tool in a file that is then processed and summarized based on the LIWC tool's dictionary. Each word in the file is searched for in the dictionary, and specific type counts are incremented based on the associated word category if found. The tool's output data include the percentage of words captured by the dictionary, standard linguistic dimensions (which include pronouns and auxiliary verbs), psychological categories and function words (e.g., cognitive, social) and personal dimensions (e.g., work, achieve, leisure and others). These different dimensions are said to capture the psychology of individuals by assessing the words they use [40] [41].

For example, consider the following sample comments:

1. "I dislike the way the customer service team works, especially the delay they cause me. This delay will no doubt affect my overall performance-appraisal when I am assessed towards the end of the year."

2. "We are aiming to have all the patches ready by the end of this release; this will provide us some space for the next one. Also, we are extremely confident that similar bug-issues will not appear in the future."

In the first comment the author is expressing dissatisfaction at the treatment received from another department, and is worried about the potential negative consequences this treatment will cause. Here the words "I", "my" and "me" are indicators of self-focus, "dislike" is associated with negative feelings, and "end" denotes some form of temporal reference. Words such as "performance-appraisal" are not captured by the LIWC dictionary, and as such the summary output for the text above would be the same whether the author was referring to achievement at sports or working on a software feature. Although these omissions may be seen as presenting a limitation, we know that the context is software development; and what is of interest here, and is being captured by the tool, is evidence of behavior.

In the second comment the author is expressing optimism that the team will succeed, and in the process finish ahead of time and with acceptable quality standards. In this quotation, the words "we" and "us" are indicators of team or collective focus, "all", "extremely" and "confident" are associated with certainty, while the words "some" and "appear" are indicators of tentative processes. As in the use of the word "performance-appraisal" in the first comment, words such as "bug-issues" and "patches" are not included in the LIWC dictionary and would not affect the context of its use – whether it was to indicate a fault in software code or a problem with one's immunity to, and treatment for, a disease. Thus, in the context of its use here to discern behaviors and attitudes, the LIWC tool's output is not adversely affected by the localized or specialized nature of software developers' vocabulary or the specific meaning with which such words are used. In addition, we also triangulated our linguistic analysis with more contextual directed CA to further support our conclusions.

In the context of software development, Rigby and Hassan [42] provide confirmation of the utility of the LIWC tool via their inspection of the Apache OSS developers' mailing list exchanges. These authors uncovered that once the top two developers signaled their intention to leave the project their communications became more negative and instructive, and they spoke mostly in the future tense and communicated with less positive emotions, when compared to their earlier communications. This study also found variations in communication behavior after releases. In studying two releases Rigby and Hassan [42] found that developers' communication was optimistic after the first release, but the opposite was evident after the second release. We also used the LIWC tool in a preliminary study of three different project areas from the Jazz repository to reveal cues for the way different teams work. In that study we found variances in behaviors among those solving different forms of software tasks [31], which led us to conjecture that the project environment may have contributed to these differences in behaviors. While we systematically examined three different project areas, probing team behaviors over different phases of their projects, and looked closely at specific team members [31], this initial study was largely exploratory and we only examined a small sample of artifacts which limited our level of inferences.

In the present work we examine core developers' behaviors along multiple linguistic dimensions. We provide a summary of the LIWC linguistic categories that were considered, along with brief theoretical justification for their inclusion, in Table 1. We consider our supplementary directed CA approach next.

## 4.4 Directed Content Analysis (CA)

The LIWC tool is applied in a top-down fashion, as its categories of language codes have been pre-determined. We anticipated that a more exploratory, bottom-up approach focused on phrases might provide different insights into core developers' interactions, behaviors and enacted roles. Such data-driven examinations have led to enhanced understanding of many issues in the software engineering and information systems domains [49]. Thus, we studied all the messages contributed in the first of the ten projects (460 messages from P1 – see Table 3) using directed CA. We employed a hybrid classification scheme adapted from related prior work [47, 50] to explore the issues underpinning our research questions. Use of a directed CA approach is appropriate when there is scope to extend or complement existing theories around a phenomenon [51], and so suited the exploration of core developers' enacted roles and possible triangulation of the findings uncovered by our linguistic analysis. The directed content analyst approaches the data analysis process using existing theories to identify key concepts and definitions as initial coding categories. In our case, we used theories examining knowledge and behaviors expressed during textual interaction [47, 50] to inform our initial categories (scales 1-8 in Table 2).

Should existing theories prove insufficient to capture all relevant insights during preliminary CA data analysis, new categories and subcategories should be created [51]. In our study, both authors of this work and two other trained coders randomly categorized 5% of the project's communications in a preliminary coding phase, and found that some aspects of Jazz team members' utterances were not captured by the first version of our CA protocol (e.g., Instructions and expressions of Gratitude were not captured). During the pilot coding exercise we also found that practitioners in Jazz communicated multiple ideas in their messages. Thus, we segmented the communication using the sentence as the unit of analysis. We extended the initial protocol by deriving new scales directly from the pilot Jazz data (contributing scales 9 to 13 in Table 2 – codes emerged in the order that they are shown in the Table), after which the first author and the two trained coders recoded all the messages. Duplicate codes were assigned to utterances that demonstrated multiple forms of collaboration, and all coding differences were discussed and resolved by consensus. We noted an 81% inter-rater agreement between the three coders using Holsti's [52] coefficient of reliability measurement (C.R). This represents excellent agreement among the coders.

**Table 1. Summary linguistic measures**

| Linguistic Category | Abbreviation | Examples | Reason for Inclusion |
|---|---|---|---|
| Pronouns | I | I, me, mine, my, I'll, I've, myself, I'm | Elevated use of first person plural pronouns (we) is evident during shared situations, whereas, relatively high use of self references (I) has been linked to individualistic attitudes [43]. Use of the second person pronoun (you) may signal the degree to which members rely on (or delegate to) other team members or their general awareness [44] of others and their activities. |
| | we | we, us, our, we've, lets, we'd, we're, we'll | |
| | you | you, your, you'll, you've, y'all, you'd, yours, you're | |
| Cognitive language | insight | think, consider, determined, idea | Software teams were previously found to be most successful when many group members were highly cognitive and were natural solution providers [45]. These traits also previously correlated with effective task analysis and brainstorming capabilities. |
| | discrep | should, prefer, needed, regardless, | |
| | tentat | maybe, perhaps, chance, hopeful | |
| | certain | definitely, always, extremely, certain | |
| Work and Achievement related language | work | feedback, goal, boss, overtime, program, delegate, duty, meeting | Individuals most concerned with task completion and achievement are said to reflect these traits during their communication. Such individuals are most concerned with task success, contributing and initiating ideas and knowledge towards task completion [46]. |
| | achieve | accomplish, attain, closure, resolve, obtain, finalize, fulfill, overcome, solve | |
| Leisure, social and positive language | leisure | club, movie, entertain, gym, party, jog, film | Individuals that are personal and social in nature are said to communicate positive emotion and social words and this trait is said to contribute towards an optimistic group climate [46-47]. Leisure related language may also be an indicator of a team friendly atmosphere. |
| | social | give, buddy, love, explain, friend, | |
| | posemo | beautiful, relax, perfect, glad, proud | |
| Negative language | negemo | afraid, bitch, hate, suck, dislike, shock, sorry, stupid, terrified | Negative emotion may affect team cohesiveness and group climate. This form of language shows discontent and resentment [48]. |

## 5. RESULTS AND ANALYSIS

Table 3 shows the ten project areas (out of 94) that were selected for analysis. As noted in Section 4.2.1, the project areas selected represented both information-rich and information-rare cases in terms of the volume of messages contributed. Projects ranged from short (59 days) to long duration (1014 days), with varying levels of communication density. The selected project artifacts amounted to 1201 software development tasks with histories, worked on by 394 contributors with profiles and responsibility information (from 474 total contributors), and 5563 messages exchanged around the 1201 tasks.

**Table 2. Coding categories for measuring interaction**

| Scale | Category | Characteristics and Example |
|---|---|---|
| 1 | Type I Question | Ask for information or requesting an answer – "Where should I start looking for the bug?" |
| 2 | Type II Question | Inquire, start a dialogue - "Shall we integrate the new feature into the current iteration, given the conflicts that were reported when we attempted same last week?" |
| 3 | Answer | Provide answer for information seeking questions - "The bug was noticed after integrating code change 305, you should start debugging here." |
| 4 | Information sharing | Share information – "Just for your information, we successfully integrated change 305 last evening." |
| 5 | Discussion | Elaborate, exchange, and express ideas or thoughts – "What was most intriguing about solving this bug is not how bugs may exist within code that went through rigorous testing... but how refactoring reveals bugs even though no functional changes are made." |
| 6 | Comment | Judgemental – "I disagree that refactoring may be considered the ultimate test of code quality." |
| 7 | Reflection | Evaluation, self-appraisal of experience – "I found solving the problems in change 305 to be exhausting, but I learnt a few techniques that should be useful in the future." |
| 8 | Scaffolding | Provide guidance and suggestions to others – "Let's document the procedures that were involved in solving this problem 305, it may be quite useful for new members joining the team in the future." |
| 9 | Instruction/ Command | Directive – "Solve task 234 in this iteration, there is quite a bit planned for the next." |
| 10 | Gratitude/ Praise | Thankful or offering commendation – "Thanks for your suggestions, your advice actually worked." |
| 11 | Off task | Communication not related to solving the task under consideration – "How was your weekend?" |
| 12 | Apology | Expressing regret or remorse – "I am very sorry for the oversight, and I am quite unhappy with the failure this has caused." |
| 13 | Not Coded | Communication that does not fit codes 1 to 12. |

## 5.1 Core Developers - Formal Roles and Task Involvement

Table 4 shows that only fourteen contributors across the ten projects met the initial density-based selection criterion for core developers (shown as bold font contributor numbers – notice that none of the members from P8 were selected initially). Thus, we followed the precedent set by Rigby and Hassan [42] and selected the top two contributors to each project, which increased the total number of core developers by six (the non-bold font contributor numbers), bringing the core developers cluster to 20. Note also from Table 4 that a few of the core developers were dominant across multiple projects (e.g., see contributors 4661 and 2419 in P1 and P2); thus, in total there were 15 distinct core developers.

We examined the recorded role information for these practitioners and found that a slight majority of the core developers were programmers (eight out of the 15 distinct cluster members), along with five team leaders and two project managers. We provide results from the mined change history data to examine core developers' involvement in software task changes in Table 5. We found an association between the level of communication and contributors' involvement in software tasks; however, this was not statistically significant (r = 0.130, n = 40, P = 0.424).

**Table 3. Summary statistics for the selected Jazz project areas**

| Project ID | WI Count | Project Area | Total Contributors | Total Messages | Period (days) |
|---|---|---|---|---|---|
| P1 | 54 | User Experience – tasks related to UI development | 33 | 460 | 304 |
| P2 | 112 | User Experience – tasks related to UI development | 47 | 975 | 630 |
| P3 | 30 | Documentation – tasks related to Web portal documentation | 29 | 158 | 59 |
| P4 | 214 | Code (Functionality) – tasks related to application development | 39 | 883 | 539 |
| P5 | 122 | Code (Functionality) – tasks related to application development | 48 | 539 | 1014 |
| P6 | 111 | Code (Functionality) – tasks related to development of application middleware | 25 | 553 | 224 |
| P7 | 91 | Code (Functionality) – tasks related to development of application middleware | 16 | 489 | 360 |
| P8 | 210 | Project Management – tasks under the project managers' control | 90 | 612 | 660 |
| P9 | 50 | Code (Functionality) – tasks related to application development | 19 | 254 | 390 |
| P10 | 207 | Code (Functionality) – tasks related to development of application middleware | 48 | 640 | 520 |
| ∑ | 1201 | - | 394 | 5563 | - |

**Table 4. Core developers**

| Project | Contributor | Density |
|---|---|---|
| P1 | **4661** | 0.85 |
|  | 2419 | 0.48 |
| P2 | **4661** | 0.74 |
|  | 2419 | 0.29 |
| P3 | **13722** | 0.50 |
|  | 4674 | 0.23 |
| P4 | **13740** | 0.40 |
|  | **11643** | 0.33 |
| P5 | **4749** | 0.45 |
|  | 4674 | 0.32 |
| P6 | **12065** | 0.74 |
|  | **13664** | 0.55 |
| P7 | **12972** | 0.80 |
|  | **13664** | 0.63 |
| P8 | 12702 | 0.28 |
|  | 2102 | 0.16 |
| P9 | **6572** | 0.58 |
|  | **12889** | 0.44 |
| P10 | **6262** | 0.61 |
|  | 13722 | 0.17 |

In Table 5 it is shown that on average more than 41% of all software tasks were initiated by the 15 core developers. These practitioners also made more than 69% of the changes to these tasks and resolved nearly 75% of all software tasks undertaken by their teams. In fact, core developers created as many as 69% of all software tasks in P6 and made 94% of changes on P9. These scores were exceeded in project P9, where core developers resolved 98% of their team's tasks. These figures are in contrast to what would be a 'typical' contribution if WIs were distributed evenly across all contributors to a project – taking this approach team members would typically have contributed to between 1.1% (for P8) and 6.3% (for P7) of their teams' WIs.

**Table 5. Activities performed by core developers**

| Project ID | % Created | % Modifications | % Resolved |
|---|---|---|---|
| P1 | 44.4 | 66.7 | 79.6 |
| P2 | 49.1 | 58.0 | 67.0 |
| P3 | 26.7 | 66.7 | 20.0 |
| P4 | 36.0 | 49.1 | 60.7 |
| P5 | 16.4 | 62.3 | 73.0 |
| P6 | 65.8 | 78.4 | 97.3 |
| P7 | 44.0 | 63.7 | 91.2 |
| P8 | 28.6 | 73.3 | 64.3 |
| P9 | 60.0 | 94.0 | 98.0 |
| P10 | 39.6 | 85.0 | 93.7 |
| **mean** | **41.1** | **69.7** | **74.5** |

## 5.2 Linguistic Analysis – Behavior Patterns

In Section 4 we introduced our procedure for selecting the core developers. We then presented a summary from our results for practitioners' communication and their involvement in task changes in Tables 4 and 5, which shows that core developers communicated the most and were also integral to their teams' actual software development portfolio. Thus, active communicators were not merely team coordinators. We also uncovered that core developers were not restricted by their formal roles, as quite often programmers leading their teams' communication worked under formal leaders. Here we extend our analysis to examine the behaviors of these core developers and compare the traits of these practitioners against those of their counterparts. We do this using an analysis of the content of the messages contributed by core developers and other practitioners using the predefined linguistic dimensions in Table 1.

We aggregated all the communication from the two groups. Those 15 practitioners classified as core developers (from the total of 394 practitioners across the ten project areas) contributed 2567 messages out of the total 5563 messages shown in Table 3. Given our sample size (both groups contributing > 2500 messages) we first evaluated the form of the data distributions by analyzing the messages in the two groups along the 13 linguistic dimensions using the Kolmogorov-Smirnov test of normality. Our data showed violations of the normality assumption, thus, we checked the paired (core and others) individual linguistic dimensions for significant differences using the non-parametric Mann-Whitney U test. We provide our results in Table 6.

Table 6 shows that core developers were less self focused, in that they used less individualistic language (e.g., I, me, my) than the other contributors, but they tended to delegate more (e.g., you, your, you're). Our Mann-Whitney U test comparing these language dimensions for the two groups confirmed that these differences were statistically significant ($p = 0.000$ and $p = 0.012$, respectively). The other team members used significantly more individualistic language, and this group also used significantly more collective language (e.g., we, our, us) ($p = 0.039$). The other team members also used significantly more insightful (e.g., think, believe, consider) ($p = 0.000$), tentative (e.g., maybe, perhaps, apparently) ($p = 0.000$) and certainty (e.g., definitely, extremely, always) ($p = 0.007$) type utterances. This pattern was the opposite for work (e.g., feedback, goal, delegate) and achievement (e.g., accomplish, attain, resolve) related language use – Table 6 shows that the core developers tended to use more work and achievement type language than the other practitioners. These findings were also statistically significant for work ($p = 0.013$), but not so for achievement language ($p = 0.063$). Of the other linguistic dimensions (leisure, social, posemo and negemo), only the leisure (e.g., club, movie, party) category produced a statistically significant finding ($p = 0.003$) in favor of the other practitioners.

We checked the ten individual projects (P1 - 10) for differences in the behavioral processes of the two groups of practitioners to ascertain if the particular project environments and/or the specific practitioners involved could have mediated the above results. We found a similar pattern of results for individualistic and delegation language across the projects; however, results for collective language were slightly different, tending to be even across the two groups. While core developers were more collective on some projects (e.g., P1 – P3, P6, P7, P9 and P10), other members were more collective on the others (e.g., P4, P5 and P8). Our Mann-Whitney U test for the individual projects (P1 - P10) for cognitive dimensions also produced a similar pattern of results as noted for the complete data set. Apart from those working on P1, the core developers for all other projects used consistently higher levels of work and achievement language ($p = 0.007$ is statistically significant for the achieve dimension in favor of the other practitioners on P1). Findings for the leisure and social dimensions were also similar to those reported for the entire data set; however, we only observed significant differences ($p = 0.017$ and $p = 0.001$) in the use of leisure utterances on P3 and P8. On the other hand, the other members involved in projects P2, P6 and P8 expressed significantly higher amounts of negative emotion ($p = 0.002$, $p = 0.002$ and $p = 0.025$ respectively).

We then checked the way the core developers expressed behaviors when they were working on more than one project. Our distributions for the selected linguistic dimensions for each of these five core developers in Table 7 were close to normal (only slightly positively skewed), so we checked for differences across three linguistic dimensions using t-tests. Table 7 shows that the core developers involved in multiple projects in our sample exhibited similar traits across those projects. Use of first-person pronouns (e.g., I, me, my) was almost identical, while there was also relative consistency in the use of social words (e.g., give, buddy, love) and discrepancy words (e.g., should, would, could).

**Table 6. Results for linguistics analysis**

| Linguistic Category | Abbrev. | Core (Mean Rank) | Others (Mean Rank) | Mann-Whitney Test ($p$-value) |
|---|---|---|---|---|
| Pronouns | I | 2711.47 | 2853.31 | 0.000 |
| | we | 2752.66 | 2818.16 | 0.039 |
| | you | 2829.86 | 2752.27 | 0.012 |
| Cognitive | insight | 2716.7 | 2848.85 | 0.000 |
| | discrep | 2779.08 | 2795.61 | 0.663 |
| | tentat | 2706.05 | 2857.94 | 0.000 |
| | certain | 2742.24 | 2827.05 | 0.007 |
| Work and Achievement | work | 2841.84 | 2742.06 | 0.013 |
| | achieve | 2828.21 | 2753.68 | 0.063 |
| Leisure, social and positive | leisure | 2738.62 | 2830.14 | 0.003 |
| | social | 2772.75 | 2801.01 | 0.490 |
| | posemo | 2748.06 | 2822.09 | 0.073 |
| Negative | negemo | 2773.92 | 2800.01 | 0.410 |

**Table 7. Results comparing differences in selected language usage for core developers involved in multiple projects**

| Contributor | Projects | t-Test: Two Sample Assuming Unequal Variance ($p$-value) | | |
|---|---|---|---|---|
| | | First-person pronouns | Social process words | Discrepancy words |
| 4661 | P1, P2 | 0.878 | 0.920 | 0.888 |
| 2419 | P1, P2 | 0.902 | 0.742 | 0.685 |
| 13722 | P3, P10 | 0.949 | 0.250 | 0.089 |
| 4674 | P3, P5 | 0.990 | 0.814 | 0.244 |
| 13664 | P6, P7 | 0.905 | 0.349 | 0.603 |

## 5.3 Directed CA - Interaction Patterns

We now take a more contextual look at the communication contributed by core and other developers when they were working on project P1. We use a directed CA approach and coded the 460 messages contributed by those involved in this project. From these messages, we recorded 1165 utterances. Figure 2 provides a summary (counts) of the interaction behavior of the core developers and those of the other contributors. In Figure 2 it is observed that core developers asked more Type I and II Questions (20 and 37 against 14 and 20 for other practitioners, respectively) and provided more Information and Instructions than the other practitioners (273 and 76 against 198 and 12, respectively). The core developers comprise only two of the 33 team members of P1 (one programmer and one team lead), however, in Figure 2 it is shown that they contributed more or similar amounts of Discussions (66 versus 63), Scaffolding (51 versus 52) and Gratitude (26 versus 25). Quite revealing was the much lower level of debate contributed by core developers when compared to the other participants (39 against 63). We performed a Chi-square test which confirmed that these differences were statistically significant ($x^2$ = 60.813, df = 12, p = 0.000). We also examined the distribution of formal roles among the 33 members of project P1 to see overall how project members contributed in their roles, and found that P1 comprised 2 'Project Managers', 11 'Team Leads' and 20 'Programmers'. Programmers, Team Leads and Project Managers contributed 47.1%, 50.5%, and 2.4% of their team's utterances, respectively.

## 6. DISCUSSION AND IMPLICATIONS

### 6.1 Discussion

*Q1. What are the core developers' enacted roles in their teams, and how are these roles occupied?* Our results show that the Jazz software practitioners that communicated the most also made the highest numbers of task changes, and were integral to their team's knowledge processes. These findings are somewhat in line with those of Cataldo et al. [12]. However, there is slight divergence between our research outcomes and those of Bird et al. [24], who found that communication increased with the need to coordinate and control, but the volume of messages an individual sent bore no relationship to the position they held in their social network. These differences may be as a result of the difficulty inherent in uniquely identifying practitioners' records in OSS repositories due to the volume of email addresses and aliases these individuals use, a problem noted by Bird et al. [24]. In Jazz however, these problems do not exist. These differences in outcomes may also reflect the challenge of studying and interpreting developers' communication processes through solely mathematical means [18]. Our findings suggest that core communicators occupied central positions in their teams, and these individuals were essential to their projects' actual development portfolio. Additionally, our findings here show that, in Jazz, core developers were not restricted by their formal role, and often times these individuals were willing to enact other roles outside of their formally assigned roles. These findings are interesting given that Jazz teams are led by a formal project manager. In Jazz a person occupying the formal 'Programmer' (contributor) role is defined as a contributor to architecture and code of a component, the 'Team Leader' (component lead) is responsible for planning and architectural integrity of the component and the 'Project Manager' (PMC) is a member of the project management committee overseeing the Jazz project. We expected that those assigned to leadership roles (and particularly project managers) would at least dominate project communication networks given their need to coordinate and manage multiple project dependencies. However, the evidence provided here is to the contrary.

Given the higher volume of messages conveyed by core developers we anticipated that these individuals would dominate knowledge sharing in their teams, and the results support this position. However, our directed CA results confirm that these individuals were integrally involved with team organization and task assignment (e.g., see measures for Answers and Instruction). It was previously established that individuals involved in such forms of (vertical) communications are seen as capable, and such individuals are often perceived by their peers as knowledge hubs, and pillars of the knowledge construction process [47, 50]. Discourses of an assertive nature (e.g., Type II Questions and Instructions) are also communicated due to a perception that little authoritative feedback is forthcoming [47], and may generally be linked to those in power. In fact, such responsibilities (and behaviors) are often associated with formal software project leadership or individuals occupying more coordination and planning related roles [45]. Core developers provided context awareness for the other team members and acted as their team's main information resource (e.g., see measures for Information sharing, Discussion and Scaffolding). Such competencies are typically associated with highly skilled roles; or those that are extremely creative, imaginative and insightful [53]. Core developers were also their teams' main implementers (see Table 5), and were team players (e.g., see their lower measures for Comment and judgmental language). Lower incidence of judgmental discourse is often required for maintaining team spirit and overcoming tension, which is critical to a positive team atmosphere [53]. While there was convergence between task changes and the volume of ideas and suggestions provided by core developers, the much lower level of judgmental attitudes expressed by these core developers suggests that these individuals also exhibited higher levels of intra-personal and inter-personal skills [54]. Our evidence suggests that the formal project managers acted as facilitators, and were happy to let their teams self-organize. Such a hands-off approach to project governance may only be feasible if team members are achievement motivated and informal leaders are present – the core developers.

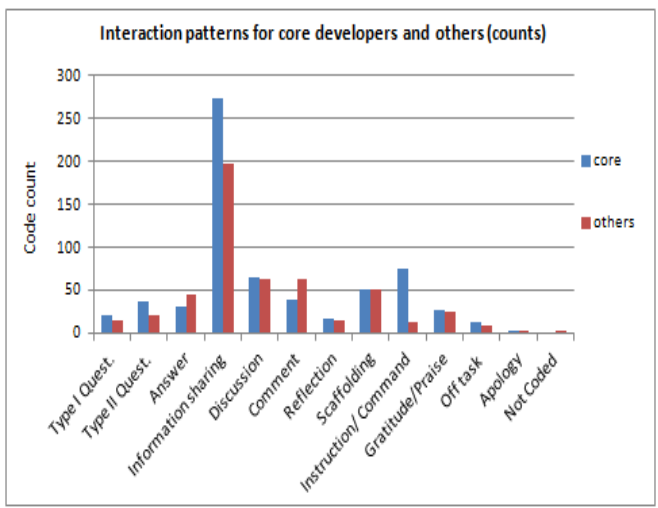

**Figure 2. Summary of project interaction for the core developers and others (for P1)**

The behaviors demonstrated by Jazz core developers may not be default behaviors, however. Such high performing members often need to possess intrinsic motivation and keen willingness to self-organize [55]. A facilitating organization and work structure are also likely to be prerequisites for encouraging high performers to work across roles as the need arises. On the basis of our results we would contend that IBM Rational is one such organization that encourages team members' performance based on their natural abilities, and that promotes non-hierarchical and informal work structures. Such configurations have long been shown to encourage tacit knowledge sharing and cross-fertilization among team members, by allowing team members to adapt and execute their tasks based on work demands [56], as was evident among Jazz practitioners. We believe this and similar environments are well suited for agile teams (particularly those working in distributed settings), and should be encouraged in order to facilitate team success.

*Q2. How do core developers' behaviors and attitudes differ from the other software practitioners?* Our linguistic findings show that, when compared to the other software practitioners, core developers were less self-focused and, although these individuals were most actively involved in task changes, they delegated more. Core developers were highly task and achievement focused, and these individuals rarely communicated off-task. These findings are triangulated by our contextual CA results. Core developers were willing to ask questions and they also provided support for others. In contrast, although our linguistic analysis shows that the other contributors tended to used more collective language, and higher amounts of cognitive language, our deeper analysis uncovered that the contributions of information, ideas and suggestions, and guidance were higher among the core developers. In fact, the other members also tended to use more judgmental language, excesses of which may lead to disharmony among teams [57]. The lower levels of this form of expression communicated by the core developers, however, may counter these higher measures for the other practitioners. Additionally, from our findings, it is clear that the core developers were the most influential in their teams.

As a group, the Jazz developers used significant amounts of positive and social language; this may have led in part to a friendly work environment, lower staff turnover and overall Jazz teams' success (over 30,000 companies are using IBM Rational Jazz tools – see jazz.net). In fact, although the Jazz teams studied here were highly social and positive, overall, it is clearly evident that these teams were also achievement focused and used significant levels of work-related language. Our more contextual analysis confirmed that Jazz teams were highly task-focused (as evidenced by the low measures for the Off-task category). Role theories have previously shown that both social and task-related behaviors are necessary to maintain team balance and team performance [46]. Social roles are said to contribute towards positive group climate, promoting harmonizing and compromising traits, while task roles are concerned with task success, contributing and initiating ideas and knowledge towards task completion. Jazz's core developers were more task focused, but these individuals were also team players. On the other hand, while the other members tended to be more individualistic and judgmental, these individuals also contributed social and positive attitudes to group climate.

## 6.2 Implications

We found that the Jazz members that communicated extensively were most integral to their teams. These individuals occupied various informal roles in their teams and were central to their teams' actual development portfolio. The other members complemented these individuals, and expressed behaviors that may be responsible for maintaining team balance. We believe that these findings have implications for software engineering research and practice.

Understanding what motivates core developers would help us to coordinate efforts aimed at detecting and molding 'software gems'. Staffing software teams with a larger ratio of these individuals – presuming they are available – may then contribute towards the reduction of incidents and project failures. However, given core developers' attitudes, their growth and performance will depend on a flexible organization culture. This is particularly necessary for distributed developments, were fluid organization roles are essential to mitigate the elevated levels of cultural and personality challenges that are likely to be introduced by distance. Organizations (and project managers) employing more rigid project governance approaches may hinder such core developers' performance, and in the process, erode the advantages associated with these members' presence. In fact, core developers such as those studied in this work may find it difficult to work under tight project controls given their naturally fluid inclination.

These findings may also have implications for tool design. Our contextual analysis shows that one half of Jazz practitioners' communications were directed at information sharing. This form of utterance, although useful for providing context awareness (i.e., explanations and information about software features, e.g., details about the outcomes of software builds), may not be as critical to the teams' development portfolios as the provision of questions, answers, suggestions and ideas. These latter, more critical types of communication may become 'lost' underneath the less significant messages (e.g., those expressing gratitude or praise). This issue was previously experienced by those involved in global software development, resulting in negative performance issues [15]. We believe that including a message tagging feature in Jazz or any similar tool (as is done for tagging software tasks) could help developers to manage this wealth of communication. During time-constrained work periods, comment tags should help practitioners to identify and consider the most critical issues first. For instance, if comment tags were labeled to express similar meanings to the categories and related scales in Table 2, a programmer coming in to work would likely review and action messages with Scales 9 (Instructions) and 1 (Questions) first, before going through the other messages in his or her order of preference.

## 7. LIMITATIONS

Although these findings are novel we acknowledge that there are shortcomings to this study that may present threats to the work's generalizability.

Measurement Validity: The LIWC language constructs used to measure team behavior have been used previously to study this subject, and were assessed for validity and reliability [40]. However, although the LIWC dictionary was able to capture around 66% of the overall words used by Jazz practitioners, the adequacy of these constructs in the specific context of software development warrants further investigation. To that end, we checked a small sample of the messages to see what might account for the remaining words being ignored by the LIWC tool and found that there were large amounts of cross referencing to other WIs in the messages, along with large amounts of highly specialized software related language (e.g., J2EE, LDAP, JACC, API, XML, TAME, JASS, Jazz, URI, REST, HTTP, Servlet, WIKI, UseCase, HTML, CVS, Dump, Config, SourceControl) evident in Jazz practitioners' exchanges. Moreover, what was of interest, and was captured by the LIWC tool, was evidence of attitude, demeanor and behavior.

Additionally, our contextual directed content analysis involving interpretation of textual data is subjective, and so questions naturally arise regarding the validity and reliability of the outcomes. In this work we employed multiple techniques to deal with these issues. First, our protocol was adapted from those previously employed and tested in the study of interaction and knowledge sharing [47, 50], and so there is a strong theoretical basis for its use. Second, we piloted the protocol and extended our instrument by deriving additional codes directly from the Jazz data, and we tested this extended protocol for accuracy, precision and objectivity, receiving an inter-rater measure indicative of excellent agreement [51].

Internal and External Validity: Although we achieved data saturation after analyzing the third project case, the tasks, history logs and messages from the ten project areas (out of 94) may not necessarily represent all the teams' processes in the repository. Additionally, work processes and work culture at IBM are also specific to that organization and may not be representative of organization dynamics elsewhere.

## 8. CONCLUSIONS AND FUTURE WORK

Given recent interest in understanding the human processes involved in contemporary software development, repository data has received increasing attention. While studies examining software repositories have provided multiple insights into the way software teams work, these investigations have approached the study of team processes using mainly mathematically-based analysis techniques. There are reservations regarding the effectiveness of these techniques in delivering understanding of the deeper psychosocial nature of team dynamics. For instance, while it has been previously uncovered that few developers dominated project communications and these individuals made most task changes during their teams' software projects, the rationale for this phenomenon has not been provided. Details around the reasons for these practitioners' distinct presence and performance, and insights into how these members contribute to team dynamics, have not been uncovered. Additionally, there are reservations over the reliability and validity of some of the software repositories that are commonly examined.

In this study we extracted and mined the IBM Rational Jazz repository, and applied deeper psycholinguistic and directed CA techniques to address these gaps. We found that Jazz's most active developers occupied various informal roles in their teams, they were central to their teams' actual development portfolios, and these practitioners exhibited behaviors that are integral for maintaining team dynamics. Additionally, we observe that as a group, Jazz developers spent most time providing context awareness for others. Given this finding, we believe a message tagging feature could reduce Jazz teams' overhead related to searching the repository of messages for the more critical questions and instructions. We believe that Jazz's core developers' performance is directly related to an organizational environment that promotes informal and organic work structures. This form of organization configuration may be necessary for agile teams, and especially for distributed developments. Our next step is to investigate how core developers' behaviors evolve over time, as we suspect that there may be some initial team arrangements that cause these developers to become hubs in their teams. Knowledge of what motivates core developers could be invaluable in coordinating efforts aimed at detecting and molding such 'software gems'.

## ACKNOWLEDGMENTS

We thank IBM for granting us access to the Jazz repository. We would also like to thank the coders for their help during our directed content analysis phase. S. Licorish is supported by an AUT VC Doctoral Scholarship Award. IBM and Jazz are trademarks of IBM Corporation.

## 10. REFERENCES

[1] Siddiqui, M. S. and Hussain, S. J. 2006. Comprehensive Software Development Model. In *Proceedings of the IEEE International Conference on Computer Systems and Applications* (March 8, 2006). 353 - 360.

[2] Boehm, B. 2006. A view of 20th and 21st century software engineering. In *Proceedings of the 28th International Conference on Software Engineering* (Shanghai, China, 2006). ACM Press, 12 - 29.

[3] Licorish, S., Philpott, A. and MacDonell, S. G. 2009 Supporting agile team composition: A prototype tool for identifying personality (In)compatibilities. In *Proceedings of the ICSE Workshop on Cooperative and Human Aspects on Software Engineering (CHASE '09)* (Vancouver, Canada, May 17, 2009). IEEE Computer Society, 66 - 73.

[4] Chin, G. 2004. *Agile Project Management: How to Succeed in the Face of Changing Project Requirements*. American Management Association, New York.

[5] El Emam, K. and Koru, A. G. 2008. A Replicated Survey of IT Software Project Failures. *IEEE Software*, 25, 5, 84-90.

[6] Standish Group. 2009. *CHAOS Summary 2009*. The Standish Group International Inc, West Yarmouth, MA.

[7] Abrahamsson, P., Marchesi, M., Succi, G., Sfetsos, P., Stamelos, I., Angelis, L. and Deligiannis, I. 2006. *Investigating the Impact of Personality Types on Communication and Collaboration-Viability in Pair Programming – An Empirical Study*. In Extreme Programming and Agile Processes in Software Engineering. Springer Berlin / Heidelberg.

[8] Rajendran, M. 2005. Analysis of team effectiveness in software development teams working on hardware and software environments using Belbin Self-Perception inventory. *Journal of Management Development*, 24, 8 (January, 2005), 738-753.

[9] Acuna, S., T, Gomez, M. and Juristo, N. 2009. How do personality, team processes and task characteristics relate to job satisfaction and software quality? *Inf. Softw. Technol.*, 51, 3, 627-639. 10.1016/j.infsof.2008.08.006.

[10] Herbsleb, J. D., Mockus, A., Finholt, T. A. and Grinter, R. E. 2001. An empirical study of global software development: distance and speed. In *Proceedings of the 23rd International Conference on Software Engineering* (Toronto, Ontario, Canada). IEEE Computer Society, 81 - 90.

[11] Shihab, E., Bettenburg, N., Adams, B. and Hassan, A. 2010. *On the Central Role of Mailing Lists in Open Source Projects: An Exploratory Study*. In New Frontiers in Artificial Intelligence. Springer Berlin / Heidelberg.

[12] Cataldo, M., Wagstrom, P., A, Herbsleb, J., D and Carley, K., M. 2006. Identification of coordination requirements: implications for the Design of collaboration and awareness tools. In *Proceedings of the 2006 20th anniversary conference on Computer Supported Cooperative Work* (Banff, Alberta, Canada). ACM, 353-362. 10.1145/1180875.1180929.

[13] Ocker, R., J and Fjermestad, J. 2008. Communication differences in virtual design teams: findings from a multi-method analysis of high and low performing experimental teams. *SIGMIS Database*, 39, 1, 51-67. 10.1145/1341971.1341977.

[14] Mistrík, I., Grundy, J., Hoek, A., Whitehead, J., Damian, D., Kwan, I. and Marczak, S. 2010. *Requirements-Driven Collaboration: Leveraging the Invisible Relationships between Requirements and People*. In Collaborative Software Engineering. Springer Berlin Heidelberg.

[15] Damian, D., Izquierdo, L., Singer, J. and Kwan, I. 2007. Awareness in the Wild: Why Communication Breakdowns Occur. In *Proceedings of the International Conference on Global Software Engineering*. IEEE Computer Society, 81-90. 10.1109/icgse.2007.13.

[16] Nguyen, T., Wolf, T. and Damian, D. 2008. Global Software Development and Delay: Does Distance Still Matter? In *Proceedings of the IEEE International Conference on Global Software Engineering, ICGSE 2008.* (17-20 Aug. 2008). IEEE Computer Society, 45-54.

[17] Abreu, R. and Premraj, R. 2009. How developer communication frequency relates to bug introducing changes. In *Proceedings of the Joint international and annual ERCIM workshops on Principles of software evolution (IWPSE) and software evolution (Evol) workshops* (Amsterdam, The Netherlands). ACM, 153-158. 10.1145/1595808.1595835.

[18] Di Penta, M. 2012. Mining developers' communication to assess software quality: Promises, challenges, perils. In *Proceedings of the 3rd International Workshop on Emerging Trends in Software Metrics (WETSoM), 2012* (Zurich, Switzerland, 3-3 June 2012). IEEE Computer Society, 1-1. 10.1109/WETSoM.2012.6226987.

[19] Abrahamsson, P., Warsta, J., Siponen, M. T. and Ronkainen, J. 2003. New directions on agile methods: a comparative analysis. In *Proceedings of the 25th International Conference on Software Engineering.* (Portland, Oregon, May, 2003). IEEE Computer Society Washington, DC, USA 244 - 254.

[20] Aranda, J. and Venolia, G. 2009. The secret life of bugs: Going past the errors and omissions in software repositories. In *Proceedings of the 31st International Conference on Software Engineering*. IEEE Computer Society, 10.1109/icse.2009.5070530.

[21] Singer, J. 1998. Practices of Software Maintenance. In *Proceedings of the Proceedings of the International Conference on Software Maintenance*. IEEE Computer Society, 139-145.

[22] Ducheneaut, N. 2005. Socialization in an Open Source Software Community: A Socio-Technical Analysis. *Comput. Supported Coop. Work*, 14, 4, 323-368. 10.1007/s10606-005-9000-1.

[23] Mockus, A., Fielding, R. T. and Herbsleb, J. D. 2002. Two case studies of open source software development: Apache and Mozilla. *ACM Trans. Softw. Eng. Methodol.*, 11, 3, 309-346. 10.1145/567793.567795.

[24] Bird, C., Gourley, A., Devanbu, P., Gertz, M. and Swaminathan, A. 2006. Mining email social networks. In *Proceedings of the 2006 international workshop on Mining Software Repositories* (Shanghai, China). ACM, 137-143. 10.1145/1137983.1138016.

[25] Shihab, E., Zhen Ming, J. and Hassan, A. E. 2009. Studying the use of developer IRC meetings in open source projects. In *Proceedings of the IEEE International Conference on Software Maintenance, 2009. ICSM 2009.* (20-26 Sept. 2009). 147-156.

[26] Wolf, T., Schroter, A., Damian, D., Panjer, L. D. and Nguyen, T. H. D. 2009. Mining Task-Based Social Networks to Explore Collaboration in Software Teams. *Software, IEEE*, 26, 1, 58-66.

[27] Wolf, T., Schroter, A., Damian, D. and Nguyen, T. 2009. Predicting build failures using social network analysis on developer communication. In *Proceedings of the 31st International Conference on Software Engineering*. IEEE Computer Society, 1-11. 10.1109/icse.2009.5070503.

[28] Aune, E., Bachmann, A., Bernstein, A., Bird, C. and Devanbu, P. 2008. Looking back on prediction: A retrospective evaluation of bug-prediction techniques. In *Proceedings of the Student Research Forum at SIGSOFT 2008/FSE 16, November 2008*.

[29] Bettenburg, N., Sascha, J., Schroter, A., Weib, C., Premraj, R. and Zimmermann, T. 2007. Quality of bug reports in Eclipse. In *Proceedings of the 2007 OOPSLA workshop on eclipse technology eXchange* (Montreal, Quebec, Canada). ACM, 21-25. 10.1145/1328279.1328284.

[30] Nagappan, N., Murphy, B. and Basili, V. 2008. The influence of organizational structure on software quality: an empirical case study. In *Proceedings of the 30th International Conference on Software Engineering* (Leipzig, Germany). ACM, 521-530. 10.1145/1368088.1368160.

[31] Licorish, S. A. and MacDonell, S. G. 2012. What Affects Team Behavior?: Preliminary Linguistic Analysis of Communications in the Jazz Repository. In *Proceedings of the ICSE Workshop on Cooperative and Human Aspects on Software Engineering (CHASE '12')* (Zurich, Switzerland, June 2, 2012). IEEE Computer Society, 83 - 89.

[32] Yin, R. 2002. *Case Study Research: Design and Methods, Third Edition, Applied Social Research Methods Series, Vol 5*. Sage Publications, Inc, Thousand Oaks, CA.

[33] Frost, R. 2007. Jazz and the Eclipse Way of Collaboration. *IEEE Softw.*, 24, 6, 114-117. 10.1109/ms.2007.170.

[34] Rich, S. 2010. IBM's jazz integration architecture: building a tools integration architecture and community

inspired by the web. In *Proceedings of the 19th International Conference on World wide web* (Raleigh, North Carolina, USA). ACM, 1379 -1382. 10.1145/1772690.1772936.

[35] Glaser, B. G. and Strauss, A. L. 1967. *The Discovery of Grounded Theory: Strategies for Qualitative Research*. Aldine Publishing Company, Chicago.

[36] Scott, J. 2000. *Social Network Analysis: A Handbook*. Sage Publications, London.

[37] Reagans, R. and Zuckerman, E. W. 2001. Networks, Diversity, and Productivity: The Social Capital of Corporate R&D Teams. *Organization Science*, 12, 4, 502-517.

[38] Zhong, X., Huang, Q., Davison, R. M., Yang, X. and Chen, H. 2012. Empowering teams through social network ties. *International Journal of Information Management*, 32, 3, 209-220. 10.1016/j.ijinfomgt.2011.11.001.

[39] Crowston, K., Wei, K., Li, Q. and Howison, J. 2006. Core and Periphery in Free/Libre and Open Source Software Team Communications. In *Proceedings of the 39th Annual Hawaii International Conference on System Sciences - Volume 06*. IEEE Computer Society, 118.111. 10.1109/hicss.2006.101.

[40] Mairesse, F., Walker, M., Mehl, M. R. and Moore, R. K. 2007. Using linguistic cues for the automatic recognition of personality in conversation and text. *J. Artif. Int. Res.*, 30, 1, 457-500.

[41] Pennebaker, J. W. and King, L. A. 1999. Linguistic Styles: Language Use as an Individual Difference. *Journal of Personality & Social Psychology*, 77, 6, 1296-1312.

[42] Rigby, P. and Hassan, A. E. 2007. What Can OSS Mailing Lists Tell Us? A Preliminary Psychometric Text Analysis of the Apache Developer Mailing List. In *Proceedings of the Fourth International Workshop on Mining Software Repositories*. IEEE Computer Society, 23-32. 10.1109/msr.2007.35.

[43] Pennebaker, J. W. and Lay, T. C. 2002. Language Use and Personality during Crises: Analyses of Mayor Rudolph Giuliani's Press Conferences. *Journal of Research in Personality*, 36, 3, 271-282. 10.1006/jrpe.2002.2349.

[44] Pennebaker, J. W., Mehl, M. R. and Niederhoffer, K. G. 2003. Psychological Aspects of Natural Language Use: Our Words, Our Selves. *Annual Review of Psychology*, 54, 1, 547-577. doi:10.1146/annurev.psych.54.101601.145041.

[45] Andre, M., Baldoquin, M. G. and Acuna, S. T. 2011. Formal model for assigning human resources to teams in software projects. *Information and Software Technology*, 53, 3, 259-275.

[46] Benne, K. D. and Sheats, P. 1948. Functional Roles of Group Members. *Journal of Social Issues*, 4, 2, 41-49.

[47] Zhu, E. 1996. Meaning Negotiation, Knowledge Construction, and Mentoring in a Distance Learning Course. In *Proceedings of the Selected Research and Development Presentations at the 1996 National Convention of the Association for Educational Communications and Technology*.

[48] Goldberg, L. R. 1981. Language and individual differences: The search for universals in personality lexicons. *Review of Personality and Social Psychology*, 2, 1, 141-165.

[49] Sheetz, S. D., Henderson, D. and Wallace, L. 2009. Understanding developer and manager perceptions of function points and source lines of code. *Journal of Systems and Software*, 82, 9, 1540-1549.

[50] Henri, F. and Kaye, A. R. 1992. *Computer conferencing and content analysis*. In Collaborative learning through computer conferencing: The Najaden papers. Springer-Verlag, New York.

[51] Hsieh, H.-F. and Shannon, S. E. 2005. Three Approaches to Qualitative Content Analysis. *Qualitative Health Research*, 15, 9 (November 1, 2005), 1277-1288. 10.1177/1049732305276687.

[52] Holsti, O. R. 1969. *Content Analysis for the Social Sciences and Humanities*, Reading, MA: Addison Wesley.

[53] Belbin, R. M. 2002. *Management teams: why they succeed or fail*. Butterworth-Heinemann, Woburn, UK.

[54] Downey, J. 2009. *Designing Job Descriptions for Software Development*. In Information Systems Development: Challenges in Practice, Theory, and Education. Springer US, USA.

[55] Moe, N. B., Dingsoyr, T. and Dyba, T. 2008. Understanding Self-Organizing Teams in Agile Software Development. In *Proceedings of the 19th Australian Conference on Software Engineering*. IEEE Computer Society, Washington, DC, USA, 76- 85.

[56] Powell, W. W. 1990. Neither market nor hierarchy: network forms of organization. In B. M. Staw and L. L. Cummints, eds. *Research in Organizational Behavior*, 12, 295-336.

[57] De Dreu, C. K. W. and Weingart, L. R. 2003. Task Versus Relationship Conflict, Team Performance, and Team Member Satisfaction: A Meta-Analysis. *Journal of Applied Psychology*, 88, 4, 741-749.